\documentclass[aps,prd,twocolumn,superscriptaddress,tightenlines,nofootinbib]{revtex4-1}
\usepackage[T1]{fontenc}
\usepackage{microtype}
\usepackage[textsize=scriptsize,backgroundcolor=red!70,linecolor=red]{todonotes}
\usepackage{booktabs}
\usepackage{dcolumn}
\usepackage{amsmath}
\usepackage{amsfonts}
\usepackage{amssymb}
\usepackage{graphicx}
\usepackage{hyperref}
\usepackage[all]{hypcap}
\usepackage{paralist}
\usepackage{multirow}

\usepackage{graphicx}
\usepackage{epstopdf}
\usepackage{epsfig}
\usepackage{dcolumn}
\usepackage{bm}
\usepackage{epsf}
\usepackage{dcolumn}
\def\be{\begin{equation}}
\def\ee{\end{equation}}
\def\bea{\begin{eqnarray}}
\def\eea{\end{eqnarray}}
\def\gsim{\ \rlap{\raise 2pt\hbox{$>$}}{\lower 2pt \hbox{$\sim$}}\ }
\def\lsim{\ \rlap{\raise 2pt\hbox{$<$}}{\lower 2pt \hbox{$\sim$}}\ }
\def\dslash{\kern-4pt \not{\hbox{\kern-2pt $\partial$}}}
\def\pslash{\not{\hbox{\kern-2pt p}}}

\def\l{{\rm L}}

\def\bra#1{\left\langle #1\right|}
\def\ket#1{\left| #1\right\rangle}

\def \be{\beta}

\def\beq{\begin{equation}}
\def\eeq{\end{equation}}
\def\bea{\begin{eqnarray}}
\def\eea{\end{eqnarray}}
\def\ber{\begin{eqnarray*}}
\def\eer{\end{eqnarray*}}
\def\bwt{\begin{widetext}}
\def\ewt{\end{widetext}}

\def\roughly#1{\mathrel{\raise.3ex\hbox
{$#1$\kern-.75em\lower1ex\hbox{$\sim$}}}}
\def\lsim{\roughly<}
\def\gsim{\roughly>}

\def\order{\lower 1.8ex \hbox{\LARGE\~{}}}

\usepackage{bm}

\def\bra#1{\left\langle #1\right|}
\def\ket#1{\left| #1\right\rangle}

\def \({\left(}
\def \){\right)}
\def \[{\left[}
\def \]{\right]}
\def \l|{\left|}
\def \r|{\right|}

\def \be{\beta}

\begin{document}
\DeclareGraphicsExtensions{.pdf,.ps}


\title{Nonstandard interactions in solar neutrino oscillations \\with Hyper-Kamiokande and JUNO}


\author{Jiajun Liao}
\affiliation{Department of Physics and Astronomy, University of Hawaii at Manoa, Honolulu, HI 96822, USA}
 
\author{Danny Marfatia}
\affiliation{Department of Physics and Astronomy, University of Hawaii at Manoa, Honolulu, HI 96822, USA}

\author{Kerry Whisnant}
\affiliation{Department of Physics and Astronomy, Iowa State University, Ames, IA 50011, USA}

\begin{abstract}

Measurements of the solar neutrino mass-squared difference from KamLAND and solar neutrino data are somewhat discrepant, perhaps due to nonstandard neutrino interactions in matter. We show that the zenith angle distribution of solar neutrinos at Hyper-Kamiokande and the energy spectrum of reactor antineutrinos at JUNO can conclusively confirm the discrepancy and detect new neutrino interactions.
  
\end{abstract}
\maketitle

\section{Introduction}
There is currently about a $2\sigma$ tension in measurements of the neutrino mass splitting, $\delta m_{21}^2\equiv m_2^2-m_1^2$, from solar and reactor neutrino experiments~\cite{Maltoni:2015kca}. The discrepancy mainly arises from the measurement of the day-night asymmetry in the Super-Kamiokande (SK) experiment~\cite{Hyper-Kamiokande:2016dsw}. The latest SK combined measurement of the day-night asymmetry is $A_{DN}^\text{SK}\equiv 2(\Phi_D-\Phi_N)/(\Phi_D+\Phi_N)=-3.3\pm1.0\pm0.5\%$~\cite{Abe:2016nxk}, where $\Phi_D$ ($\Phi_N$) is the measured solar neutrino flux during the day (night).{\footnote{SNO, on the other hand, does not disfavor a vanishing day-night asymmetry at more than $2\sigma$~\cite{Aharmim:2011vm}. } This day-night asymmetry is extracted for $\delta m_{21}^2=4.8\times 10^{-5}$ $\text{eV}^2$, while the global best-fit value (which is dominated by KamLAND data~\cite{Gando:2010aa}) is $\delta m_{21}^2$=$7.5\times 10^{-5}$ $\text{eV}^2$~\cite{Maltoni:2015kca}, for which the day-night asymmetry is $-1.7\%$.


Analyzing the SK data in Table XII of Ref.~\cite{Abe:2016nxk}, and the KamLAND data in Fig. 1 of Ref.~\cite{Gando:2010aa} in the standard model (SM), we find the preferred parameters shown in Fig.~\ref{fig:SK} (which are consistent with those shown in Fig. 2 of Ref.~\cite{Maltoni:2015kca}). From Fig.~\ref{fig:SK}, we see a tension between the allowed regions at SK and KamLAND. For $\sin^2\theta_{12}$ close to the global best fit value of 0.31, SK data prefer a smaller value of $\delta m_{21}^2$ than KamLAND. However, due to large uncertainties at SK, as shown in Fig.~\ref{fig:SK}, current experiments are not able to resolve the tension. Nevertheless, if this discrepancy is due to a new physical effect, future solar and reactor experiments that have better control of the systematic uncertainties and larger datasets will see a significant difference in their measurements of $\delta m_{21}^2$, and could provide conclusive evidence for the existence of new physics. 
\begin{figure}
\centering
\includegraphics[width=0.4\textwidth]{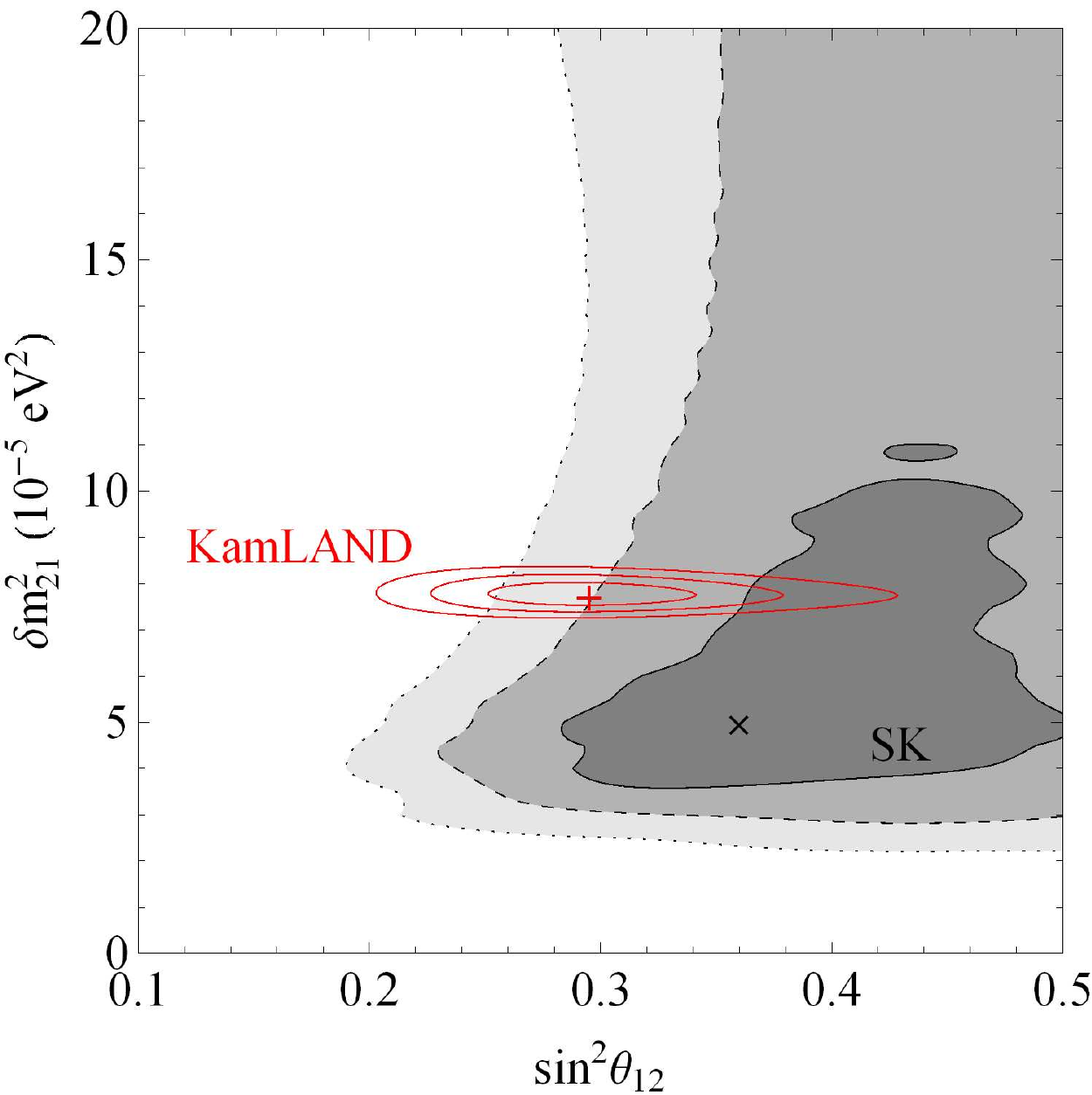}
\caption{$1\sigma$, $2\sigma$ and $3\sigma$ allowed regions from SK (shaded regions) and KamLAND (ellipses) data in the SM. The cross (plus) sign marks the best-fit values at SK (KamLAND). We fix $\sin^2\theta_{13}=0.023$, $\sin^2\theta_{23}=0.43$ and
$\delta m_{31}^2=2.43\times 10^{-3} \text{eV}^2$.
}
\label{fig:SK}
\end{figure}

In this work, we explore this tension in the measurement of $\delta m_{21}^2$ with future solar neutrino data at Hyper-Kamiokande (HK)~\cite{Hyper-Kamiokande:2016dsw} and future reactor antineutrino data at JUNO~\cite{An:2015jdp}. We use the framework of nonstandard
interactions (NSI), which provides a model-independent way of studying new physics in neutrino
oscillation experiments; for reviews see Ref.~\cite{Ohlsson:2012kf}. In particular, we focus on the day-night asymmetry and the zenith-angle distribution in solar neutrino experiments in the presence of matter NSI. Matter NSI can be described by
dimension-six four-fermion operators of the
form~\cite{Wolfenstein:1977ue, Guzzo:1991hi}
\bea
  \label{eq:NSI}
  \mathcal{L}_\text{NSI} = - 2\sqrt{2}G_F
   \epsilon^{fC}_{\alpha\beta} 
        \left[ \overline{\nu}_\alpha \gamma^{\rho} P_L \nu_\beta \right] 
        \left[ \bar{f} \gamma_{\rho} P_C f \right] + \text{h.c.}\,,
\eea
where $\alpha, \beta=e, \mu, \tau$, $C=L,R$, $f=u,d,e$, and
$\epsilon^{fC}_{\alpha\beta}$ specifies the
strength of the new interaction in units of $G_F$.

The JUNO experiment will measure $\delta m_{21}^2$ to the percent level~\cite{An:2015jdp}, but it is not sensitive to the NSI parameters due to its short baseline and low neutrino energy.  The HK solar neutrino oscillation probabilities are strongly dependent on the NSI parameters due to the MSW effect~\cite{Wolfenstein:1977ue,Mikheev:1986gs}, but HK will not precisely measure $\delta m_{21}^2$ due to systematic uncertainties. A combination of the two experiments could provide direct evidence for the existence of new physics if the $\delta m_{21}^2$ discrepancy persists in
HK and JUNO data.

The paper is organized as follows. In Section~II, we analyze the day-night asymmetry with NSI. In Section~III, we describe our simulations of HK and JUNO. We discuss our results in Section~IV, and sum up in Section~V.

\section{Day-night asymmetry with NSI}
\subsection{Formalism}
The Hamiltonian in the three neutrino framework for neutrino propagation in the presence of matter NSI can be written in the flavor basis as
\bea
H =   U\text{diag}
\left(0, \frac{\delta m^2_{21}}{2E_\nu} , \frac{\delta m^2_{31}}{2E_\nu}\right)
U^\dagger + V\,,
\eea
where $U$ is the Pontecorvo-Maki-Nakagawa-Sakata mixing matrix~\cite{Agashe:2014kda},
\bea
&&U = R_{23}\Gamma_\delta R_{13}\Gamma_\delta^\dagger R_{12} =
\\\nonumber
&&\left(\begin{array}{ccc}
c_{13} c_{12} & c_{13} s_{12} & s_{13} e^{-i\delta}
\\
-s_{12} c_{23} - c_{12} s_{23} s_{13} e^{i\delta} &
c_{12} c_{23} - s_{12} s_{23} s_{13} e^{i\delta} &
c_{13} s_{23}
\\
s_{12} s_{23} - c_{12} c_{23} s_{13} e^{i\delta} &
-c_{12} s_{23} - s_{12} c_{23} s_{13} e^{i\delta} &
c_{13} c_{23}
\end{array} \right)
\eea
where $R_{ij}$ represents a real rotation by an angle $\theta_{ij}$ in the $ij$ plane, $\Gamma_\delta=\text{diag}(1,1,e^{i\delta})$, and $s_{ij}$ and $c_{ij}$ denote $\sin \theta_{ij}$ and $\cos\theta_{ij}$ respectively.  The potential $V$ originating from interactions of neutrinos in matter is
\bea
V =  \sqrt2 G_F N_e\left(\begin{array}{ccc}
1 + \epsilon_{ee} & \epsilon_{e\mu} & \epsilon_{e\tau}
\\
\epsilon_{e\mu}^* & \epsilon_{\mu\mu} & \epsilon_{\mu\tau}
\\
\epsilon_{e\tau}^*& \epsilon_{\mu\tau}^* & \epsilon_{\tau\tau}
\end{array}\right)\,,
\label{eq:V}
\eea
where
$\epsilon_{\alpha\beta}\equiv\sum\limits_{f}\epsilon^{f}_{\alpha\beta}\frac{N_f}{N_e}$ with
$
\epsilon_{\alpha\beta}^f \equiv
\sum\limits_{C} \epsilon^{fC}_{\alpha\beta}\,
$, and $N_{f}$ is the number density of fermion $f$ at a given location.

Following Ref.~\cite{Akhmedov:2004rq}, we work in the new basis $\ket{\tilde{\nu}}=\tilde{U}^\dagger\ket{\nu_\alpha}$, with $\tilde{U}=R_{23}\Gamma_\delta R_{13}$. The Hamiltonian in the new basis becomes
\bea
&&\tilde{H}=R_{12}\text{diag}
\left(0, \frac{\delta m^2_{21}}{2E_\nu} , \frac{\delta m^2_{31}}{2E_\nu}\right)R_{12}^T+\tilde{V}\,,
\\\nonumber
&=&\left(\begin{array}{ccc}
s_{12}^2\frac{\delta m^2_{21}}{2E_\nu}+ \tilde{V}_{11} & s_{12}c_{12}\frac{\delta m^2_{21}}{2E_\nu}+ \tilde{V}_{12} & \tilde{V}_{13}
\\
s_{12}c_{12}\frac{\delta m^2_{21}}{2E_\nu}+ \tilde{V}_{12}^* & c_{12}^2\frac{\delta m^2_{21}}{2E_\nu}+ \tilde{V}_{22} &  \tilde{V}_{23}
\\
 \tilde{V}_{13}^* &  \tilde{V}_{23}^* & \frac{\delta m^2_{31}}{2E_\nu}+ \tilde{V}_{33}
\end{array}\right)\,,
\eea
where $\tilde{V}=\tilde{U}^\dagger V \tilde{U}$. For solar neutrinos, since $\frac{|\delta m^2_{31}|}{2E_\nu}\gg\tilde{V}_{ij}$, the third mass eigenstate decouples from the other mass eigenstates, and the evolution is governed by an effective $2\times 2$ submatrix.
After subtracting a constant diagonal matrix from the Hamiltonian, the effective matrix can be written as
\bea
\label{eq:Heff}
&\tilde{H}_\text{eff}&=\frac{\delta m_{21}^2}{4E_\nu}\times
\\\nonumber
&&\left(\begin{array}{cc}
-\cos2\theta_{12}+2\hat{A}(c_{13}^2-\epsilon_D) & \sin2\theta_{12}+2\hat{A}\epsilon_N
\\
\sin2\theta_{12}+2\hat{A}\epsilon_N^* & \cos2\theta_{12}+2\hat{A}\epsilon_D
\end{array}\right)\,,
\eea
where $\hat{A}=2\sqrt{2}G_FN_eE_\nu/\delta m_{21}^2$ and $\epsilon_X=\sum_f N_f \epsilon_X^f/N_e$ ($X=N,D$) with~\cite{Gonzalez-Garcia:2013usa}
\bea
\label{eq:epsD}
    \epsilon_D^f &=&
        -\frac{c_{13}^2}{2} \big( \epsilon_{ee}^f - \epsilon_{\mu\mu}^f \big)
        + \frac{s_{23}^2 - s_{13}^2 c_{23}^2}{2}
        \big( \epsilon_{\tau\tau}^f - \epsilon_{\mu\mu}^f \big) 
    \nonumber\\& &
    + \text{Re}\left[ c_{13} s_{13}e^{i\delta} \big( s_{23} \, \epsilon_{e\mu}^f
      + c_{23} \, \epsilon_{e\tau}^f \big)\right. 
    \nonumber\\
    &&\qquad\left.- \big( 1 + s_{13}^2 \big) c_{23} s_{23}  \epsilon_{\mu\tau}^f \right]
\,,
\eea
\bea
\label{eq:epsN}
  \epsilon_N^f &=&
  c_{13} \big( c_{23} \, \epsilon_{e\mu}^f - s_{23} \, \epsilon_{e\tau}^f \big)
  \\\nonumber
  &+& s_{13} e^{-i\delta} \left[
    s_{23}^2 \, \epsilon_{\mu\tau}^f - c_{23}^2 \, \epsilon_{\mu\tau}^{f*}
    + c_{23} s_{23} \big( \epsilon_{\tau\tau}^f - \epsilon_{\mu\mu}^f \big)
    \right]\,.
\eea
Solar neutrinos produced in the core of the Sun arrive at the surface of the Earth as an incoherent sum of the three mass-eigenstates $\nu_1$, $\nu_2$ and $\nu_3$. During the day, the neutrinos only travel a few km in the Earth, and the $\nu_e$ survival probability from the source to the detector can be written as~\cite{Akhmedov:2004rq}
\bea
\label{eq:PD}
P_D = P_{e1}^S P_{1e}^V + P_{e2}^S P_{2e}^V + P_{e3}^S P_{3e}^V\,,
\eea
where the superscript $V$ represents neutrinos propagating in vacuum, and we have
\bea
P_{1e}^V =c_{13}^2c_{12}^2\,,\qquad P_{2e}^V = c_{13}^2s_{12}^2\,, \qquad P_{3e}^V =s_{13}^2\,.
\eea

The superscript $S$ represents neutrinos traveling in the Sun. We checked that the neutrino propagation in the Sun with NSI can be treated as adiabatic for the parameters we consider in this paper. For adiabatic propagation, we have 
\begin{eqnarray}
P_{e1}^S &=& c_{13}^2\cos^2 \theta_{12}^m\,,\nonumber
\\
P_{e2}^S &=& c_{13}^2\sin^2 \theta_{12}^m\,, 
\\
P_{e3}^S &=& s_{13}^2\,,\nonumber
\end{eqnarray}
where $\theta_{12}^m$ can be found by diagonalizing the Hamiltonian in Eq.~(\ref{eq:Heff}) at the production point, i.e.,
\bea
\tan 2\theta_{12}^m=\frac{|\sin2\theta_{12}+2\hat{A}_S\epsilon_N^S|}{\cos2\theta_{12}-\hat{A}_S(c_{13}^2-2\epsilon_D^S)}\,.
\eea
Here $\hat{A}_S=2\sqrt{2}G_FN_e^SE_\nu/\delta m_{21}^2$, and $\epsilon_X^S=\sum_f N_f^S \epsilon_X^f/N_e^S$ with $N_{f}^S$ being the number density of fermion $f$ at the production point.

During the night, the neutrinos travel a large distance through the Earth, and the survival probability becomes 
\bea
\label{eq:PN}
P_N = P_{e1}^S P_{1e}^E + P_{e2}^S P_{2e}^E + P_{e3}^S P_{3e}^E\,,
\eea
where the superscript $E$ represents neutrino propagation in the Earth. Since the third mass eigenstate decouples from the other mass eigenstates, $P_{3e}^E =s_{13}^2$. Also, from probability conservation, $P_{1e}^E =c_{13}^2-P_{2e}^E$. However, the calculation of $P_{2e}^E$ in the presence of NSI is nontrivial. Here we derive a simple expression of $P_{2e}^E$ for a constant density profile. 

For neutrino evolution in Earth matter, due to the decoupling of the third eigenstate, the effective Hamiltonian has the same form of Eq.~(\ref{eq:Heff}) with $\hat{A}$ and $\epsilon_X$ replaced by $\hat{A}_E=2\sqrt{2}G_FN_e^EE_\nu/\delta m_{21}^2$ and $\epsilon_X^E=\sum_f N_f^E \epsilon_X^f/N_e^E$, respectively, where $N_{f}^E$ is the number density of fermion $f$ in the Earth. Then the effective Hamiltonian can be diagonalized by~\cite{Liao:2015rma}
\bea
U'=\left(\begin{array}{cc}
\cos\tilde{\theta} & \sin\tilde{\theta} e^{-i\phi} 
\\
-\sin\tilde{\theta} e^{i\phi} & \cos\tilde{\theta} 
\end{array}\right)\,,
\eea
where
\bea
\tan2\tilde{\theta}=\frac{|\sin2\theta_{12}+2\hat{A}_E\epsilon_N^E|}{\cos2\theta_{12}-\hat{A}_E(c_{13}^2-2\epsilon_D^E)}\,,
\eea
and
\bea
\phi=-\text{Arg}\left(\sin2\theta_{12}+2\hat{A}_E\epsilon_N^E\right)\,.
\eea
Then the evolution matrix in the new basis can be written as~\cite{Akhmedov:2004rq}
\bea
\tilde{S}=\left(\begin{array}{ccc}
\tilde{\alpha} & \tilde{\beta} & 0
\\
-\tilde{\beta}^* & \tilde{\alpha}^* & 0
\\
0 & 0 & \tilde{\gamma}
\end{array}\right)\,.
\eea
For a constant density profile, 
\bea
\,\,\,\,\,\,\,\,\,\tilde{\alpha}&=&\cos \omega L + i \cos2\tilde{\theta} \sin \omega L\,,
\nonumber
\eea
\bea
\label{eq:abc}
\tilde{\beta}&=&- i \sin 2\tilde{\theta} e^{-i\phi} \sin \omega L\,,
\eea
\bea
\tilde{\gamma}&=& \exp(-i\frac{\delta m_{31}^2L}{2E_\nu})\,,\,\,\quad
\nonumber
\eea
where 
\bea
&\omega&=\frac{\delta m^2_{21}}{4E_\nu}\times
\\\nonumber
&&\sqrt{(\cos2\theta_{12}-\hat{A}_E(c_{13}^2-2\epsilon_D^E))^2+|\sin2\theta_{12}+2\hat{A}_E\epsilon_N^E|^2}\,.
\eea

The evolution matrix in the neutrino flavor basis becomes
\bea
S=\tilde{U}\tilde{S}\tilde{U}^\dagger\,,
\eea
and we have 
\bea
P_{2e}^E &=& |\bra{\nu_e}(SU)\ket{\nu_2}|^2
\\\nonumber
&=&c_{13}^2\left[s_{12}^2|\tilde{\alpha}|^2+c_{12}^2|\tilde{\beta}|^2+\sin2\theta_{12}\text{Re}(\tilde{\alpha}^*\tilde{\beta})\right]\,.
\eea
Plugging the expressions in Eq.~(\ref{eq:abc}) into the above equation, we get
\bea
P_{2e}^E &=& c_{13}^2\sin2\tilde{\theta}\sin^2\omega L\times
\\\nonumber
&&\left(\cos2\theta_{12}\sin2\tilde{\theta}-\sin2\theta_{12}\cos2\tilde{\theta}\cos\phi\right)
\\\nonumber
&-&c_{13}^2\sin2\theta_{12}\sin2\tilde{\theta}\sin\phi\sin\omega L\cos\omega L+c_{13}^2s_{12}^2\,.
\eea
For earth matter, $\hat{A}_E\ll 1$, so we expand the above equation to leading order in
$\hat{A}_E$ and find
\bea
P_{2e}^E &\approx& \hat{A}_E c_{13}^2 \sin2\theta_{12}\sin^2 \frac{\delta m_{21}^2L}{4E_\nu}\times
\\\nonumber
&&\left[2\cos 2\theta_{12}\text{Re}(\epsilon_N^E)+\sin2\theta_{12} (c_{13}^2-2\epsilon_D^E)\right]
\\\nonumber
&+&\hat{A}_Ec_{13}^2\sin 2\theta_{12}\sin\frac{\delta m_{21}^2L}{2E_\nu}\text{Im}(\epsilon_N^E)+c_{13}^2s_{12}^2\,.
\eea
Then from Eqs.~(\ref{eq:PD}) and~(\ref{eq:PN}), the day-night symmetry is
\bea
A_{DN}&\equiv&\frac{2(P_D-P_N)}{P_D+P_N}\nonumber
\\
&\approx&\frac{2\hat{A}_Ec_{13}^4\sin2\theta_{12}\cos2\theta_{12}^m }{c_{13}^4(1+\cos2\theta_{12}\cos2\theta_{12}^m)+2s_{13}^4} \times
\\\nonumber
&\qquad& \left[\sin^2 \frac{\delta m_{21}^2L}{4E_\nu}\left(2\cos 2\theta_{12}\text{Re}(\epsilon_N^E)\right.\right.
\\\nonumber
&&\left.\left.+\sin2\theta_{12} (c_{13}^2-2\epsilon_D^E)\right)+\sin\frac{\delta m_{21}^2L}{2E_\nu}\text{Im}(\epsilon_N^E)\right]
\,.
\eea
Since the day-night asymmetry is generally measured by integrating over the zenith angle and the oscillations in the above equation are averaged out, we obtain
\bea
\label{eq:Adn}
\left\langle A_{DN}\right\rangle(E_\nu) &\approx&\frac{\hat{A}_Ec_{13}^4\sin2\theta_{12}\cos2\theta_{12}^m }{c_{13}^4(1+\cos2\theta_{12}\cos2\theta_{12}^m)+2s_{13}^4}\times
\\\nonumber
&&\left[2\cos 2\theta_{12}\text{Re}(\epsilon_N^E)+\sin2\theta_{12} (c_{13}^2-2\epsilon_D^E)\right]\,.
\eea
We have checked that the above equation is consistent with Eq. (13) in Ref.~\cite{Friedland:2004pp} for the two-flavor case.

\subsection{Numerical analysis}
Using Eq.~(\ref{eq:Adn}), we estimate the day-night asymmetry for different SM and NSI parameters. 
In Fig~\ref{fig:Adn-SM}, we show iso-$\left\langle A_{DN}\right\rangle$ contours in the $\sin^2\theta_{12}-\delta m_{21}^2$ plane in the SM. We fix $\sin^2\theta_{13}=0.023$, $E_\nu=7.0$ MeV and Earth density $\rho_E=3.0 \text{ g}/\text{cm}^3$. The day-night asymmetry depends strongly on $\delta m_{21}^2$, and its size decreases as $\sin^2\theta_{12}$ increases in the second octant. In particular, a smaller value of $\delta m_{21}^2$ yields a larger $|\left\langle A_{DN}\right\rangle|$.
\begin{figure}
\centering
\includegraphics[width=0.4\textwidth]{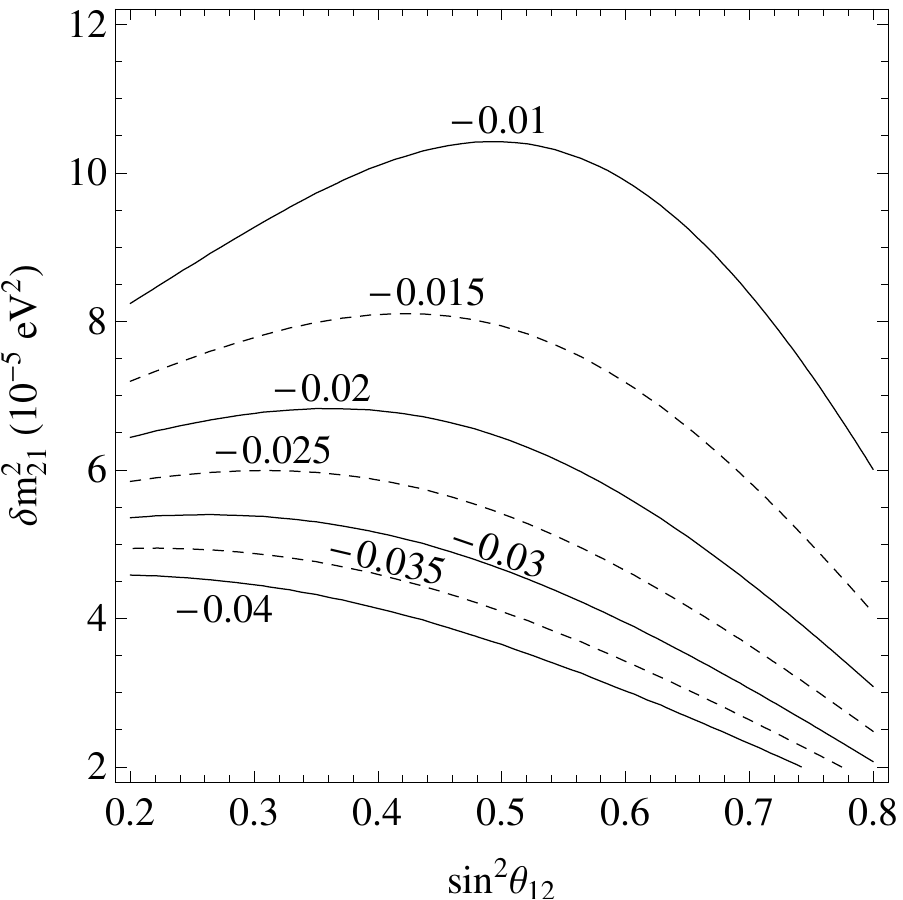}
\caption{Iso-$ \left\langle A_{DN}\right\rangle $ contours in the $\sin^2\theta_{12}-\delta m_{21}^2$ plane in the SM. Here $\sin^2\theta_{13}=0.023$, $E_\nu=7.0$ MeV and $\rho_E=3.0 \text{ g}/\text{cm}^3$.
}
\label{fig:Adn-SM}
\end{figure}

Now we study the dependence of the day-night asymmetry on the NSI parameters. For the NSI parameters, we assume there are no nonstandard couplings to electrons since they would affect the electron-neutrino scattering cross section, yielding NSI at the SK and HK detectors.
We also assume the NSI couplings to the up and down quarks are the same for simplicity. 

We first examine the dependence on the diagonal NSI parameters. We consider the case in which only $\epsilon_{ee}^u=\epsilon_{ee}^d$ is nonzero. From Eqs.~(\ref{eq:epsD}) and~(\ref{eq:epsN}), we see that $\epsilon_N^E=0$ and $\epsilon_D^E$ is linearly dependent on $\epsilon_{ee}^u$. We show the iso-$\left\langle A_{DN}\right\rangle $ contours in the space of $\delta m_{21}^2$ and $\epsilon_{ee}^u$ in Fig.~\ref{fig:Adn-NSIee}. We find that for fixed $\delta m_{21}^2$, a more positive value of $\epsilon_{ee}^u$ implies a larger $|\left\langle A_{DN}\right\rangle|$. This result can be understood from Eq.~(\ref{eq:Adn}). Since the dominant contribution to the change in $\left\langle A_{DN}\right\rangle $ comes from the factor $\left[2\cos 2\theta_{12}\text{Re}(\epsilon_N^E)+\sin 2\theta_{12} (c_{13}^2-2\epsilon_D^E)\right]$, and $\epsilon_D^E$ is proportional to $-\epsilon_{ee}^u$, as $\epsilon_{ee}^u$ increases, $|\left\langle A_{DN}\right\rangle|$ becomes larger.
This yields a degeneracy between $\delta m_{21}^2$ and $\epsilon_{ee}^u$ in the measurement of the day-night asymmetry, i.e., a day-night asymmetry that is consistent with $\delta m_{21}^2=4.8\times 10^{-5}$ $\text{eV}^2 $ in the SM can also be obtained with $\delta m_{21}^2=7.5\times 10^{-5}$ $\text{eV}^2 $ and $\epsilon_{ee}^u=\epsilon_{ee}^d\sim 0.1$. 
\begin{figure}
\centering
\includegraphics[width=0.4\textwidth]{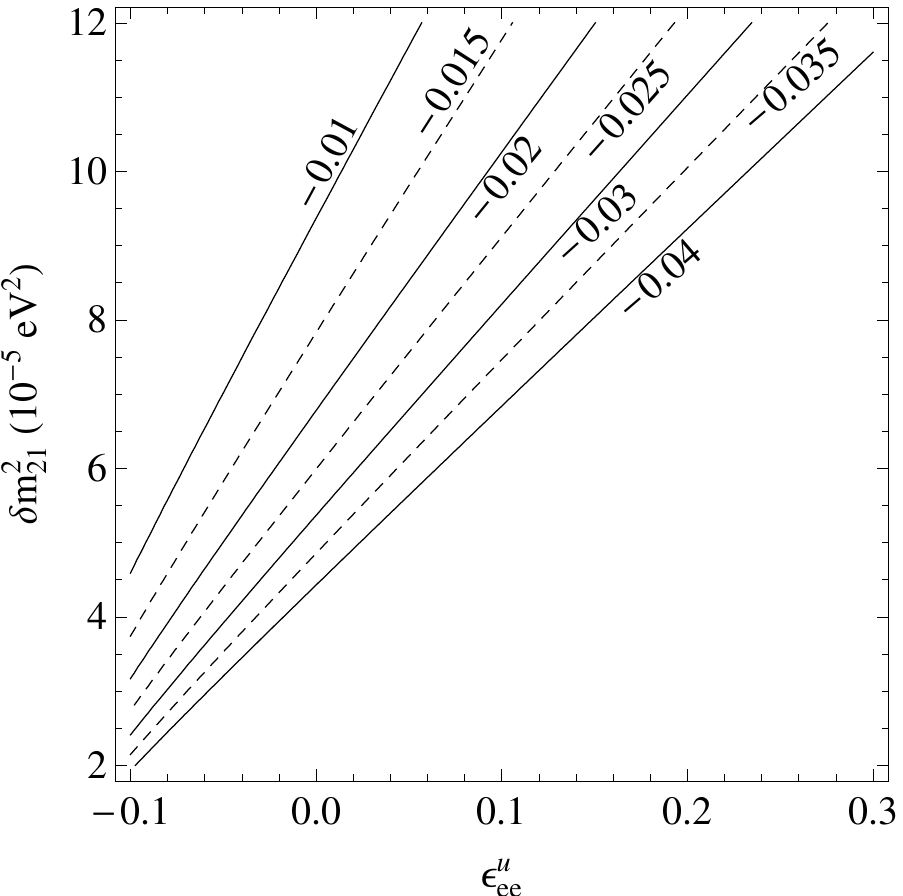}
\caption{Iso-$\left\langle A_{DN}\right\rangle$ contours in the $\epsilon_{ee}^u-\delta m_{21}^2$ plane. The parameters are the same as in Fig.~\ref{fig:Adn-SM}, and $\sin^2\theta_{12}=0.031$. For the NSI parameters we assume $\epsilon_{ee}^u=\epsilon_{ee}^d$ and all other NSI parameters are zero. 
}
\label{fig:Adn-NSIee}
\end{figure}

We also checked the dependence of the day-night asymmetry on the off-diagonal NSI parameters. Here we consider the case in which only $\epsilon_{e\tau}^u=\epsilon_{e\tau}^d$ is nonzero. As can be seen from Eqs.~(\ref{eq:epsD}) and~(\ref{eq:epsN}), $\epsilon_D^E$ is suppressed by $\sin\theta_{13}$ and $\epsilon_N^E$ is proportional to $-\epsilon_{e\tau}^u$ in this case. We first assume $\delta=0$ and $\epsilon_{e\tau}^u$ is real for simplicity. 
In Fig.~\ref{fig:Adn-NSIet}, we show the iso-$\left\langle A_{DN}\right\rangle$ contours in the space of $\delta m_{21}^2$ and $\epsilon_{e\tau}^u$.  The results in Fig.~\ref{fig:Adn-NSIet} can be understood from Eq.~(\ref{eq:Adn}). As $\epsilon_{e\tau}^u$ approaches 0.2, the factor $\left[2\cos 2\theta_{12}\text{Re}(\epsilon_N^E)+\sin 2\theta_{12} (c_{13}^2-2\epsilon_D^E)\right]$ approaches zero. We also checked the complex case by varying $\delta$ and the phase of $\epsilon_{e\tau}^u$, and find that for $\delta m_{21}^2$=$7.5\times 10^{-5}$ $\text{eV}^2$, the day-night asymmetry is always smaller than 2\% for $|\epsilon_{e\tau}^u|<0.4$. Since from Eq.~(\ref{eq:epsN}) we know that the dominant contribution to $\epsilon_N^f$ comes from $\epsilon_{e\mu}^f$ and $\epsilon_{e\tau}^f$, and the global-fit constraints on $\epsilon_{e\mu}^f$ are stronger than on $\epsilon_{e\tau}^f$~\cite{Gonzalez-Garcia:2013usa}, an off-diagonal NSI parameter always gives a small day-night asymmetry for $\delta m_{21}^2$=$7.5\times 10^{-5}$ $\text{eV}^2$. We henceforth focus on the diagonal NSI parameters.
\begin{figure}
\centering
\includegraphics[width=0.4\textwidth]{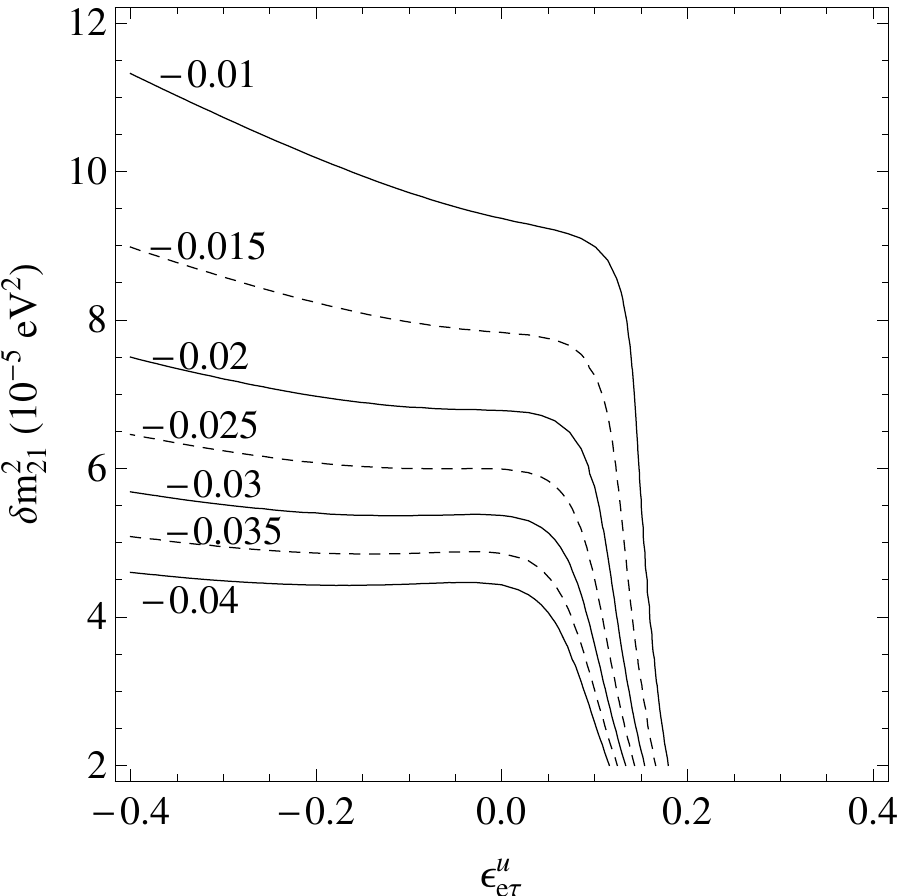}
\caption{Iso-$ \left\langle A_{DN}\right\rangle $ contours in the $\epsilon_{e\tau}^u-\delta m_{21}^2$ plane. The parameters are the same as in Fig.~\ref{fig:Adn-SM}, and $\sin^2\theta_{12}=0.031$, $\sin^2\theta_{23}=0.43$ and $\delta=0$. For the NSI parameters, we assume $\epsilon_{e\tau}^u=\epsilon_{e\tau}^d$ is real, and all other NSI parameters are zero.
}
\label{fig:Adn-NSIet}
\end{figure}

\section{Experimental simulations}
\subsection{Hyper-Kamiokande}
Solar neutrino experiments like HK detect neutrinos via the elastic scattering reaction,
\bea
\nu_x+e^-\rightarrow \nu_x'+e^-\,,
\eea
The expected event rate for the reconstructed electron kinetic energy of $T$ is~\cite{Barger:2001bw}
\bea
R(T)&=& \mathcal{N}\int dE_\nu \times
\\\nonumber
&&\left[\Phi_B(E_\nu)+1.462\times 10^{-3}\Phi_\text{hep}(E_\nu)\right]S^{D,N}(E_\nu)\,,
\eea
where $\mathcal{N}$ is the overall normalization that gives the expected event rate in the absence of oscillations, $\Phi_B$ ($\Phi_\text{hep}$) is the normalized $^8B$ (hep) neutrino flux, and the factor $1.462\times 10^{-3}$ is the relative total flux of hep to $^8B$ neutrinos in the standard solar model (SSM) (B16-GS98)\cite{Vinyoles:2016djt}. The effective cross section is 
\bea
S^{D,N}(E_\nu)= P^{D,N}\sigma_e+(1-P^{D,N})\sigma_\mu\,,
\eea
with 
\bea
\sigma_i=\int dT \int dT' \frac{d\sigma_i}{dT'}(E_\nu, T')g(T,T')\,,
\eea
where $i=e,\mu$, $T'$ is the true electron kinetic energy, $\frac{d\sigma_i}{dT'}(E_\nu, T')$ is the differential scattering cross section with radiative corrections taken from Ref.~\cite{Bahcall:1995mm}, and the energy resolution $g(T,T')$ is given by 
\bea
g(T,T')=\frac{1}{\sqrt{2\pi}\sigma(T')}\text{exp}\left[-\frac{(T-T')^2}{2\sigma(T')^2}\right]\,.
\eea
Since an energy resolution of 10\% at 10 MeV is achievable at the HK experiment~\cite{energy-resolution}, we choose the energy resolution function, 
\bea
\sigma(T') = (0.316\, {\rm{MeV}})\sqrt{\frac{T'}{\rm{MeV}}} \,.
\eea

Due to Earth matter effects, the electron neutrino survival probability at night is zenith angle dependent. Given a particular zenith angle, the relative amount of time that the detector is exposed to the Sun is determined by the latitude of the detector site. We use the exposure function at Kamiokande from Ref.~\cite{Bahcall:1997jc}, and weight each zenith angle by the exposure function. To obtain the survival probabilities, we adopt the average value of the production-point densities of the electron, up-quark and down-quark in the Sun from Ref.~\cite{Bahcall:2000nu}, and use the GLoBES software~\cite{GLOBES} with the new physics tools developed in Ref.~\cite{Kopp:2007ne} to calculate $P_{2e}^E$ numerically.

We first simulate the detector with a fiducial volume of 0.56 Mt and the electron kinetic energy threshold of 7.0 MeV from the old HK design~\cite{Abe:2011ts}. We normalize the number of events in our simulation to 200 events per day~\cite{Abe:2011ts}. Since the HK collaboration has updated their design with a new two-tank configuration for the detector~\cite{Hyper-Kamiokande:2016dsw}, we change the fiducial volume to 0.187 Mt per tank and assume the threshold energy is 5.0 MeV. Ergo, we expect 152 events per day per tank for the new design. The 2TankHK-staged configuration~\cite{Hyper-Kamiokande:2016dsw} has one tank taking data for 6 years and a second tank is added for another 4 years. We checked that our sensitivity to the day-night asymmetry is consistent with Fig. 134 in Ref.~\cite{Hyper-Kamiokande:2016dsw} for a 6.5 MeV energy threshold.

In our simulation of the HK experiment, we also consider two Earth density profiles: the Preliminary Reference Earth Model (PREM)~\cite{Dziewonski:1981xy} and the FLATCORE model~\cite{flatcore}, in which the density of the core is a constant, as shown in Fig.~\ref{fig:profile}. Note that the FLATCORE model does not match the Earth's mass, and we only use it as an example to study the effects of the Earth's density profile on our results.
\begin{figure}
\centering
\includegraphics[width=0.4\textwidth]{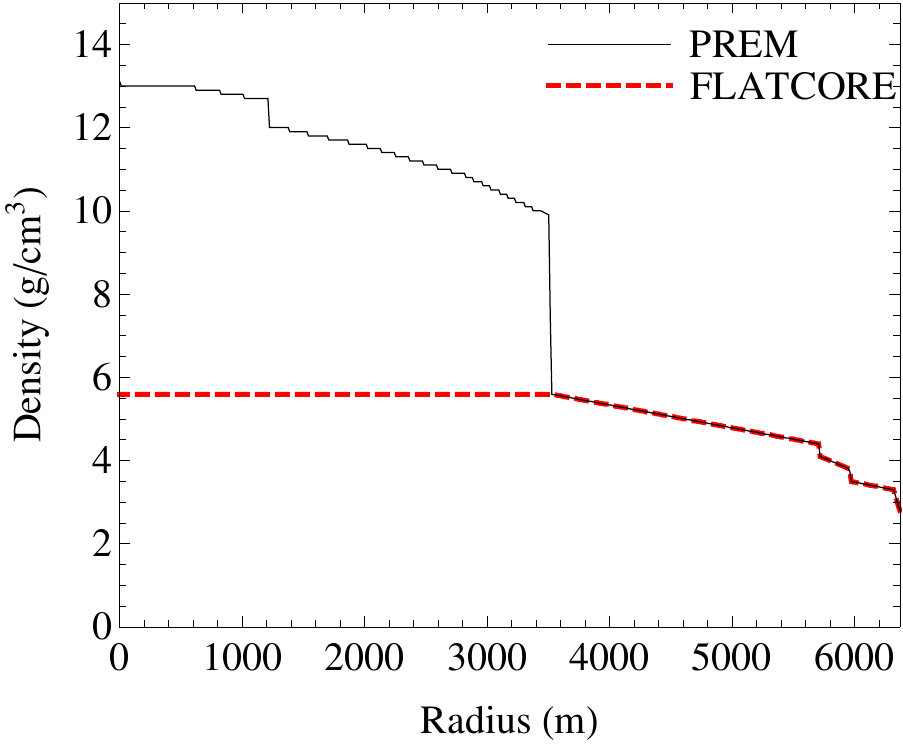}
\caption{The density profile of the Earth in two models.
}
\label{fig:profile}
\end{figure}

\subsection{JUNO}
The 20-kt liquid scintillator JUNO experiment will detect reactor antineutrinos from two reactor complexes with a total power of 36 GW via the inverse beta-decay reaction,
\bea
\bar{\nu}_e+p\rightarrow e^++n\,.
\eea
%
%
Besides the primary goal of determining the neutrino mass hierarchy, JUNO will also provide a precise measurement of the solar neutrino oscillation parameters. We simulate the JUNO experiment using the GLoBES software with the tools developed in Refs.~\cite{Kopp:2007ne, Kopp:2008ds}. The baseline of the experiment is 52.5 km and we take the detector energy resolution to be $3\%/\sqrt{E(\text{MeV})}$. With 6 years running, the detector will collect a total of $1.52\times 10^5$ events. An overall normalization error of 5\% and a linear energy scale uncertainty of 3\% is implemented in our simulation~\cite{Blennow:2013vta}. We consider 200 bins from 1.8 MeV to 8.0 MeV, and checked that the spectrum produced from our simulation is in good agreement with that in Fig. 2-15 of Ref.~\cite{An:2015jdp}.

\section{Results}

\subsection{Resolving the tension in $\delta m_{21}^2$}
Since solar data are not sensitive to parameters related to $m_3$ or the $CP$ phase for the case
we are considering, we fix $\sin^2\theta_{13}=0.023$, $ \sin^2\theta_{23}=0.43 $ and $\delta=0$.
We simulate HK and JUNO data with \mbox{$\delta m_{21}^2=7.5\times 10^{-5}$ $\text{eV}^2$}, $\sin^2\theta_{12}=0.31$,  $\delta m_{31}^2=2.43\times 10^{-3} \text{ eV}^2$ for the normal mass hierarchy (NH), and $\epsilon_{ee}^u=\epsilon_{ee}^d=0.1$, which gives a prediction for the day-night asymmetry that agrees with the current measurement at SK. 
We first perform a fit to only the SM parameters for each experiment separately to show how parameter degeneracies can occur with nonzero NSI, then perform a fit to the NSI parameters for the two experiments combined to study their ability to reject the SM. We always marginalize over the normal and inverted mass hierarchy (IH).
 

\subsubsection{Day-night asymmetry}
As an example, we first only use the day-night asymmetry in the HK analysis. The experimentally 
 measured day-night asymmetry is defined as 
\bea
A_\text{DN}^\text{exp}\equiv \frac{2(N_D-N_N)}{N_D+N_N}\,,
\eea
where $N_D$ ($N_N$) denotes the total number of events detected in the day (night) time. We fit the SM to the simulated data with NSI. From Fig.~\ref{fig:dna}, we see an allowed region for HK around $\delta m_{21}^2 =4.8 \times 10^{-5}$ $\text{eV}^2$. Note that the dependence of the HK allowed regions on $\sin^2\theta_{12}$ is consistent with the prediction of Eq.~(\ref{eq:Adn}) shown in Fig.~\ref{fig:Adn-NSIee}. The two sets of allowed regions for JUNO (shown for comparison) around $\sin^2\theta_{12}=0.31$ and $0.69$ are a consequence of the generalized mass-hierarchy degeneracy~\cite{Bakhti:2013ora}. Although the exact generalized mass-hierarchy degeneracy requires $\epsilon_{ee}\rightarrow -\epsilon_{ee}-2$, since JUNO is not sensitive to the NSI parameters, an approximate degeneracy holds. 

\begin{figure}
\centering
\includegraphics[width=0.4\textwidth]{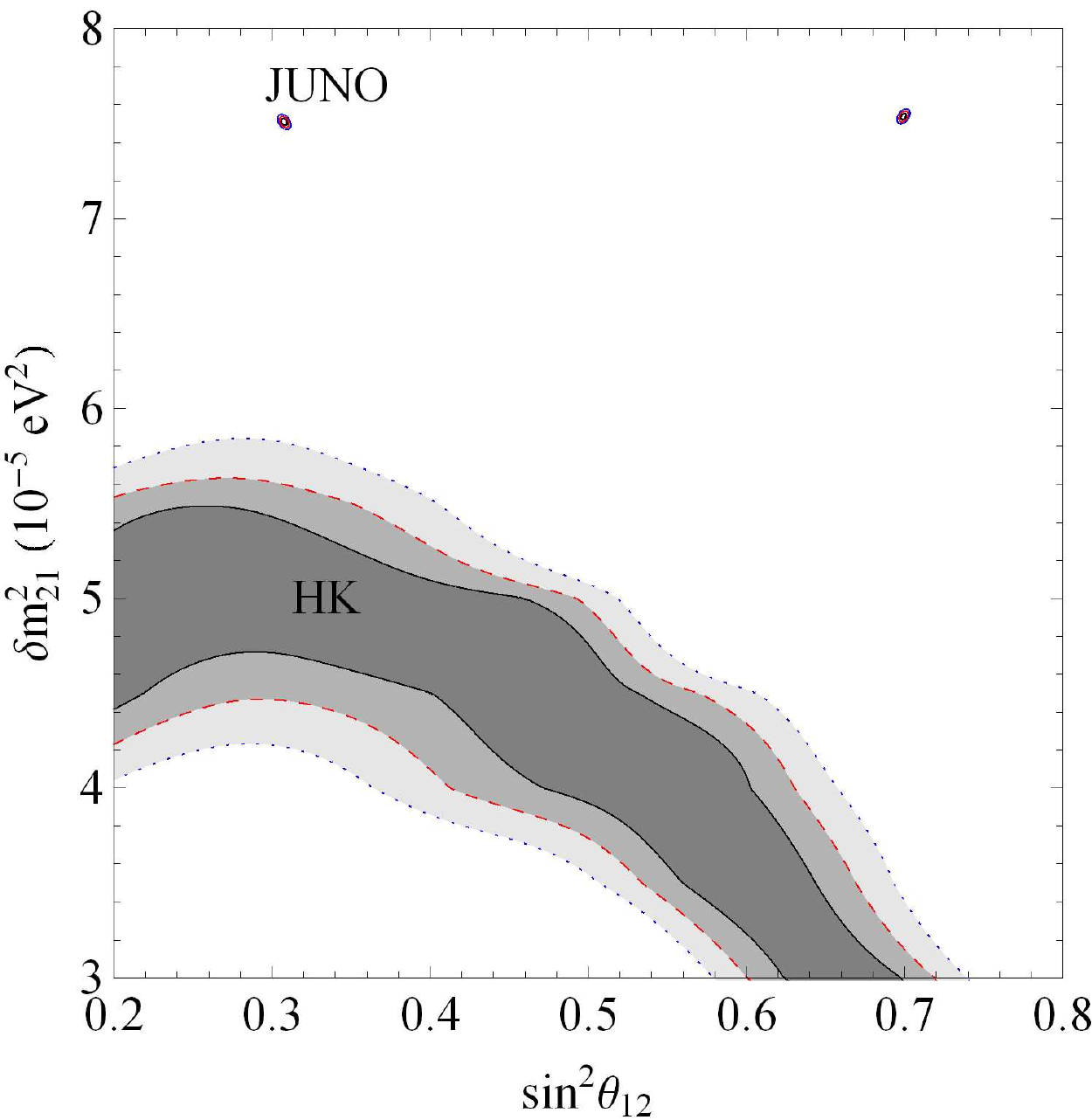}
\caption{$1\sigma$, $2\sigma$ and $3\sigma$ allowed regions for HK and JUNO; for JUNO, note the second set of allowed regions at $\sin^2\theta_{12}=0.69$. The data are simulated with $\delta m_{21}^2=7.5\times 10^{-5}$ $\text{eV}^2$, $\sin^2\theta_{12}=0.31$, NH,  $\epsilon_{ee}^u=\epsilon_{ee}^d=0.1$, and fit with the SM allowing for both mass hierarchies. Here only the day-night asymmetry is used in the analysis, and the PREM model is used for the Earth density profile.
}
\label{fig:dna}
\end{figure}

\subsubsection{Zenith-angle distribution}
We now consider one bin with daytime data and six equisized (in the cosine of the zenith angle)
nighttime bins, and define
\bea
\chi^2_{HK} = \sum_{i=1}^{7} \frac{(\alpha N_{i}^\text{fit} - N_{i}^\text{data})^2}{N_{i}^\text{data}}+\frac{(1-\alpha)^2}{\sigma_\alpha^2}\,,
\eea
where $\sigma_\alpha=12\%$ is the flux uncertainty in the SSM (B16-GS98)\cite{Vinyoles:2016djt}.
The results of a SM parameter space scan are shown in Fig.~\ref{fig:zenith}. The allowed regions around $\delta m_{21}^2 =4.8 \times 10^{-5}$ $\text{eV}^2$ persist. Compared to Fig.~\ref{fig:dna}, we see that the new analysis gives a better constraint on $\sin^2\theta_{12}$; however, the allowed regions in $\delta m_{21}^2$ are similar to those obtained  from the day-night asymmetry. Hence the sensitivity to $\delta m_{21}^2$ at HK mainly comes from the day-night asymmetry, and HK alone cannot distinguish between the SM and NSI scenarios.

\begin{figure}
\centering
\includegraphics[width=0.4\textwidth]{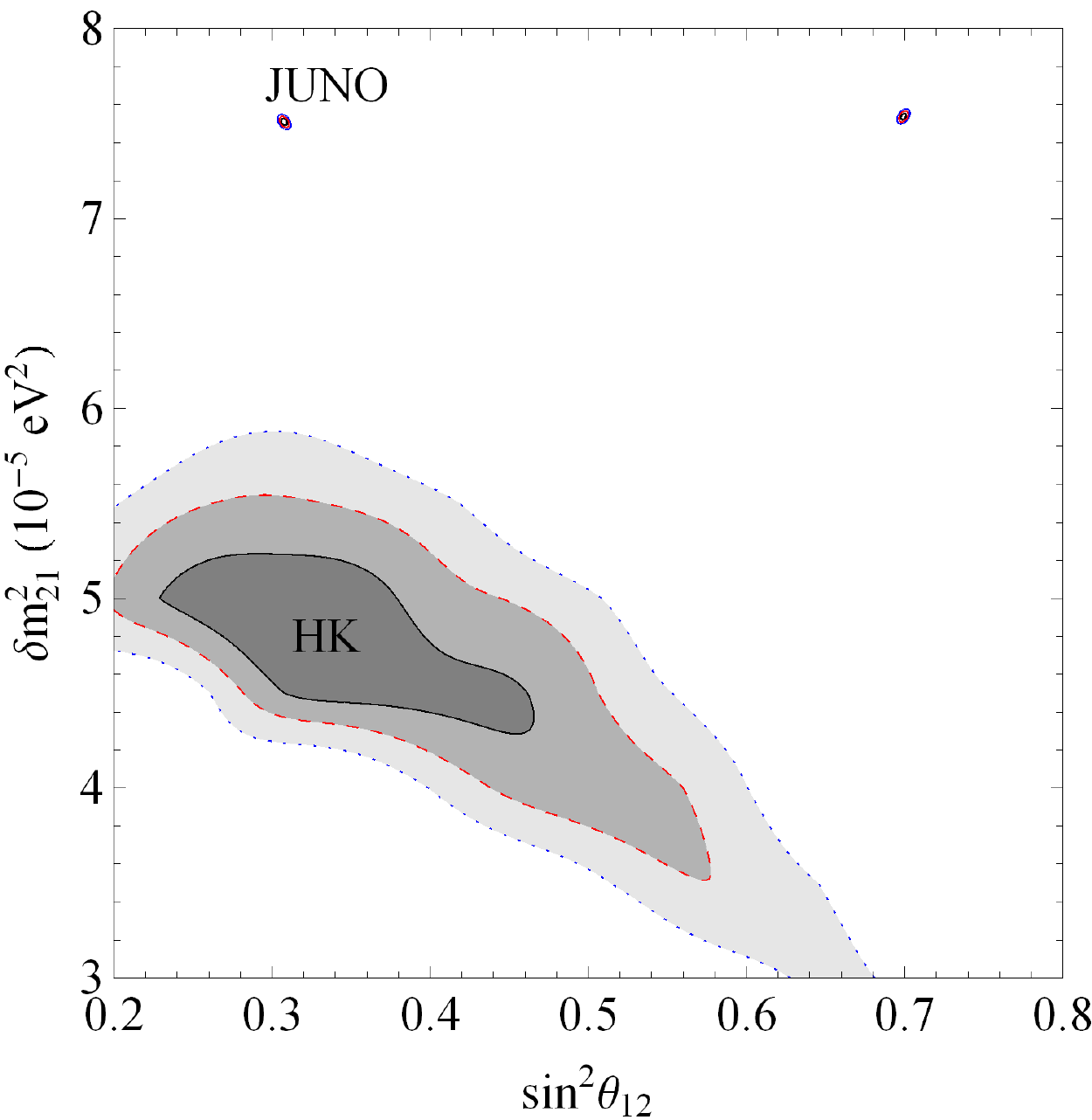}
\caption{
Same as Fig.~\ref{fig:dna}, except that the zenith-angle distribution is used in the analysis.
}
\label{fig:zenith}
\end{figure}

We also simulated data using the FLATCORE model for the Earth density profile. Then we fit the SM assuming the PREM model for the Earth density profile. The best-fit $\chi^2$ to the HK data is 2.3, indicating that the PREM model provides a good fit to data simulated with the FLATCORE model. The allowed regions shown in Fig.~\ref{fig:prem} are similar to those in Fig.~\ref{fig:zenith}. 

\begin{figure}
\centering
\includegraphics[width=0.4\textwidth]{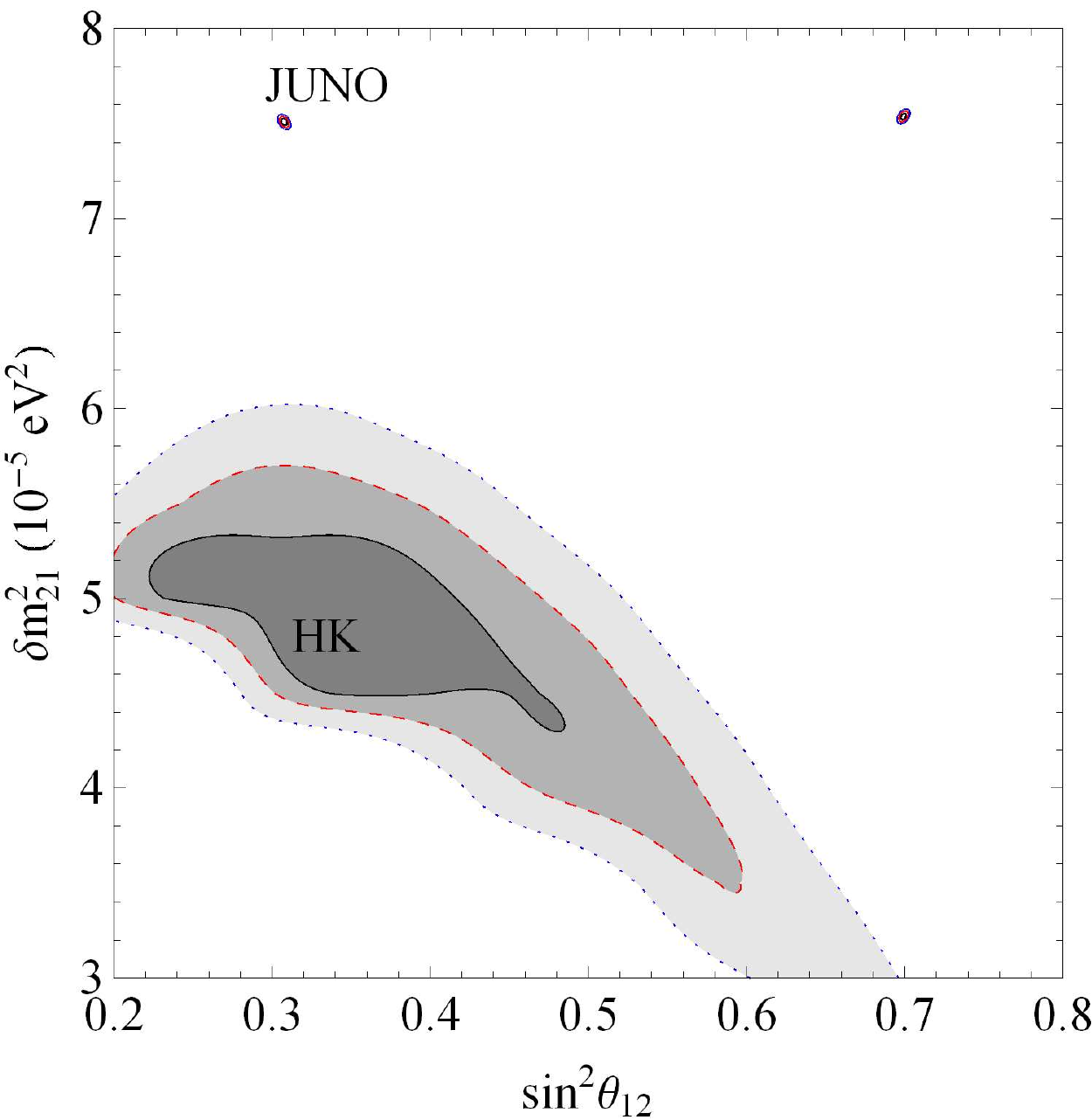}
\caption{Same as Fig.~\ref{fig:zenith}, except that the data are simulated with the FLATCORE model and fit with the PREM model.
}
\label{fig:prem}
\end{figure}

\subsubsection{HK and JUNO combined analysis}
Although HK data alone cannot distinguish between the SM and NSI scenarios, when combined with reactor data, measurements of the NSI parameters may be possible. We combine the data from JUNO and HK, and study their sensitivities to the NSI parameters. For the HK analysis, we use the zenith-angle distribution. Using simulated data with $\epsilon_{ee}^u=\epsilon_{ee}^d=0.1$ at JUNO and HK, we scan over the range of $\epsilon_{ee}^u$ that is consistent with the global fit in Ref.~\cite{Gonzalez-Garcia:2013usa}. After marginalizing over $\delta m_{21}^2$, $\theta_{12}$ and the mass hierarchy, we plot $\sqrt{\Delta\chi^2}$ as a function of $\epsilon_{ee}^u$ for the JUNO and HK combined analysis. The solid curves in Fig.~\ref{fig:eps-p1} show that the SM (with $\epsilon_{ee}^u=0$) is ruled out at 7.6$\sigma$, but 
large negative values of $\epsilon_{ee}^u$ are allowed at less than 3$\sigma$ due to the generalized mass-hierarchy degeneracy. 

\begin{figure}
\centering
\includegraphics[width=0.4\textwidth]{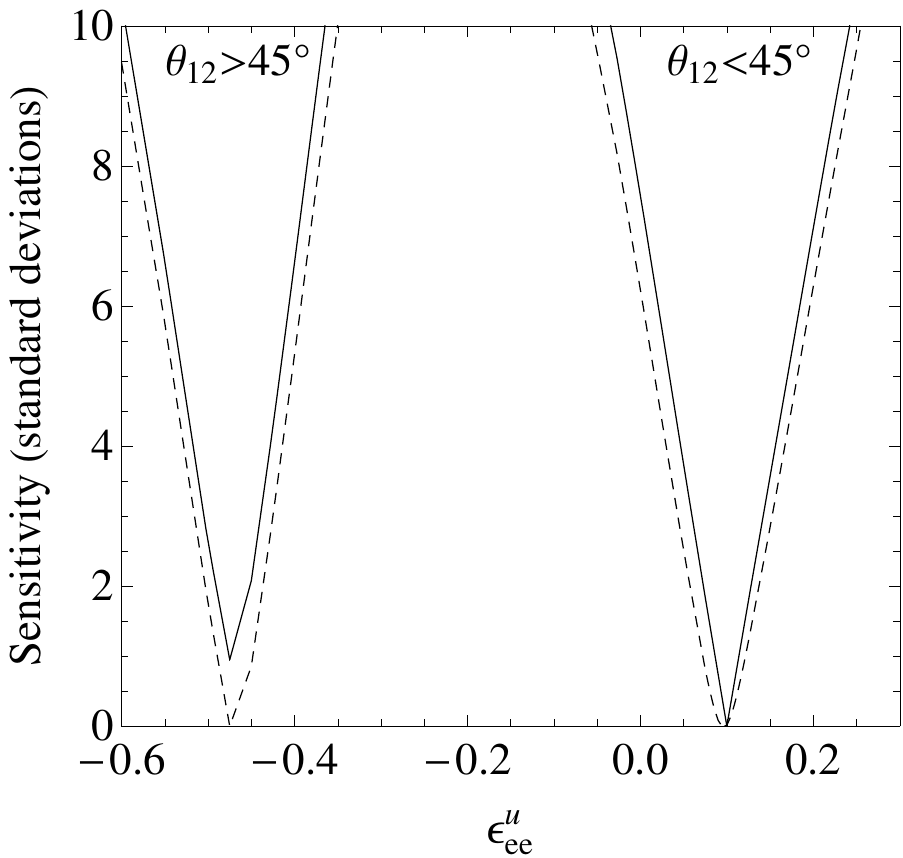}
\caption{The sensitivity to $\epsilon_{ee}^u$ for the HK and JUNO combined analysis when the true $\epsilon_{ee}^u=0.1$. The data are simulated with $\delta m_{21}^2=7.5\times 10^{-5}$ $\text{eV}^2$, $\sin^2\theta_{12}=0.31$ and NH. We assume $\epsilon_{ee}^u=\epsilon_{ee}^d$, and all other NSI parameters are zero. The solid (dashed) curves correspond to the case in which the data are simulated with the PREM (FLATCORE) model for the Earth density profile and fit with the PREM model. 
}
\label{fig:eps-p1}
\end{figure}

In order to test the effect of the Earth density profile on our results, we also simulate the data with the FLATCORE model, and fit the data assuming the PREM model. The results are shown in Fig.~\ref{fig:eps-p1} as the dashed curves. As expected, the sensitivity is reduced if the Earth density profile employed is inaccurate. However, the SM is still excluded at 6.2$\sigma$.

\subsection{Detecting NSI}
We now study the significance with which $\epsilon_{ee}^u=\epsilon_{ee}^d\neq 0$ can be established by ruling out the SM.
We simulate data with $\delta m_{21}^2=7.5\times 10^{-5}$ $\text{eV}^2$, NH, $\sin^2\theta_{12}=0.31$
 ($\sin^2\theta_{12} = 0.7)$,  and values of $\epsilon_{ee}^u$ that are roughly consistent with the global fit in Ref.~\cite{Gonzalez-Garcia:2013usa} for the first (second) octant of $\theta_{12}$. 
For each value of $\epsilon_{ee}^u$ we calculate the sensitivity to reject the SM allowing for both mass hierarchies. From Figs.~\ref{fig:scan1} and~\ref{fig:scan2}, we see that the combination of HK and JUNO data can exclude the SM at high confidence for a range of $\epsilon_{ee}^u$ values.{\footnote{Guided by the generalized mass-hierarchy degeneracy, we also simulated data for the IH by fixing $\delta m_{31}^2=-2.355\times 10^{-3} \text{ eV}^2$, thus yielding $(\delta m_{32}^2)_\text{IH}=-(\delta m_{31}^2)_\text{NH}$ for the simulated data. The sensitivity to reject the SM is identical to that in Figs.~\ref{fig:scan1} and~\ref{fig:scan2}.}  The kink on the left side of the curve in Fig.~\ref{fig:scan1} arises because the second octant of $\theta_{12}$ and IH provides a better fit than the first octant for $\epsilon_{ee}^u\sim -0.05$.
If $\sin^2\theta_{12}=0.7$, Fig.~\ref{fig:scan2} shows that for $\epsilon_{ee}^u \sim -0.4$, the SM is allowed at less than 3$\sigma$ as a result of the generalized mass-hierarchy degeneracy. 

\begin{figure}
\centering
\includegraphics[width=0.4\textwidth]{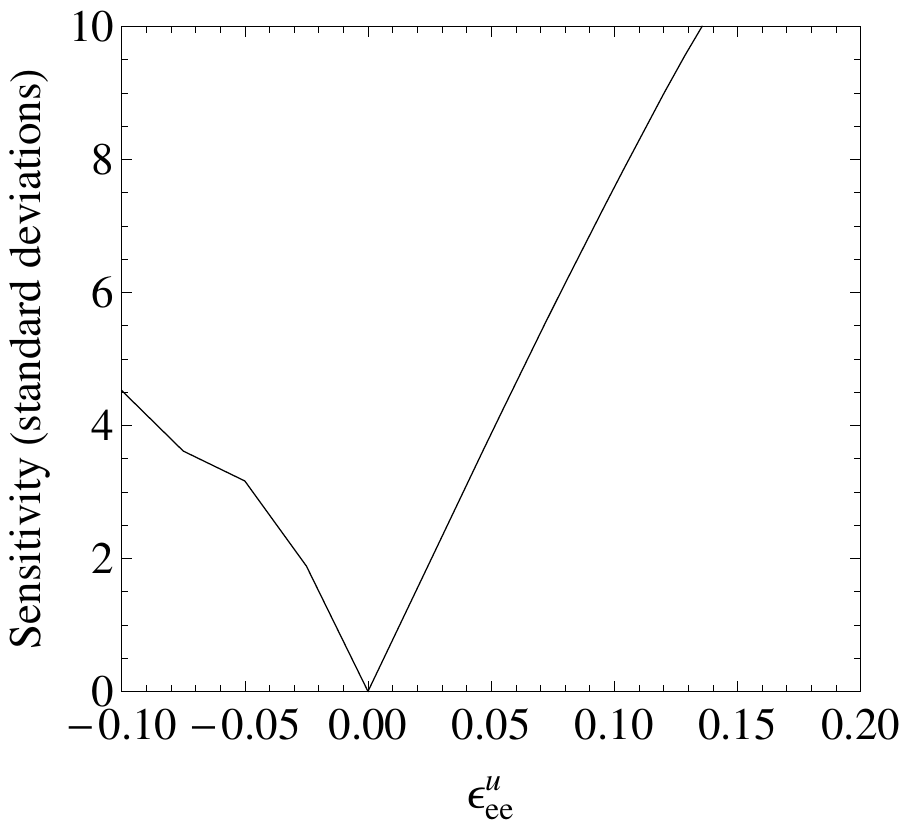}
\caption{The sensitivity to reject the SM as a function of true $\epsilon_{ee}^u$ for the HK and JUNO combined analysis. The data are simulated with $\delta m_{21}^2=7.5\times 10^{-5}$ $\text{eV}^2$, $\sin^2\theta_{12}=0.31$, and NH.
}
\label{fig:scan1}
\end{figure}

\begin{figure}
\centering
\includegraphics[width=0.4\textwidth]{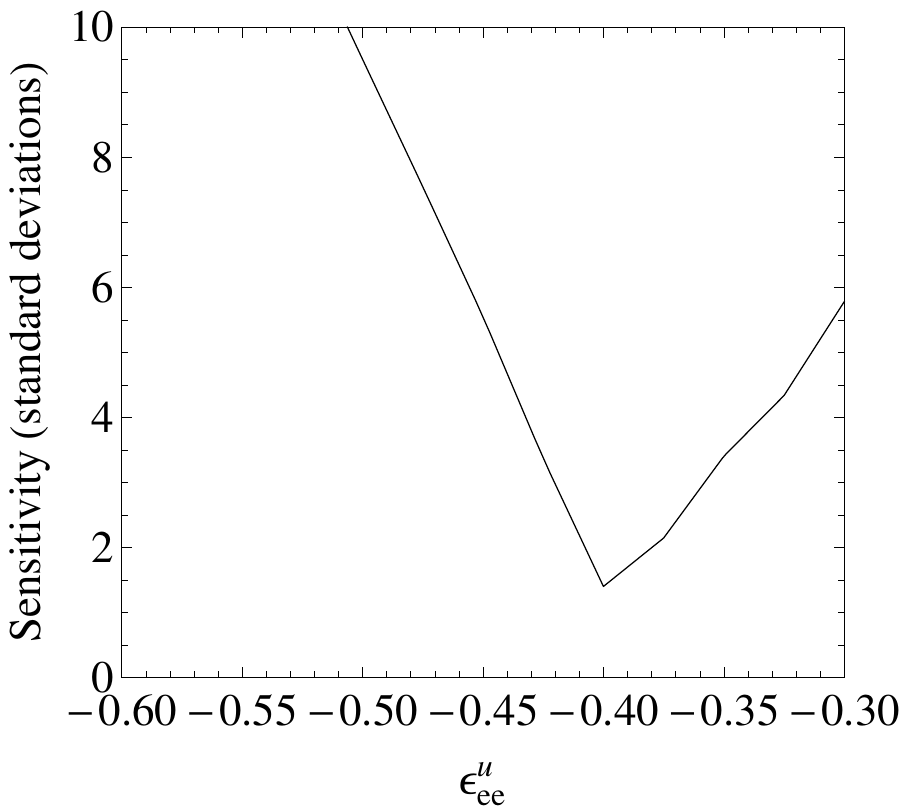}
\caption{Same as Fig.~\ref{fig:scan1}, except that the data are simulated with $\sin^2\theta_{12}=0.70$. 
}
\label{fig:scan2}
\end{figure}

\section{Summary}
We explored the discrepancy in the current measurements of $\delta m_{21}^2$ from the SK solar neutrino and KamLAND reactor antineutrino experiments in the framework of NSI. Since the discrepancy mainly stems from the measurement of the day-night asymmetry, we first derived an analytic formula for the day-night asymmetry in the presence of NSI in the three-neutrino framework. We studied the dependence of the day-night asymmetry on both the diagonal and off-diagonal NSI parameters using the formula. We find that a diagonal NSI parameter could yield a large day-night asymmetry. In particular, for $\delta m_{21}^2=7.5\times 10^{-5}$ $\text{eV}^2$, the value preferred by KamLAND, $\epsilon_{ee}^u=\epsilon_{ee}^d=0.1$ could give a day-night asymmetry that agrees with the current measurement at SK. We also find that an off-diagonal NSI parameter always yields a small day-night asymmetry for $\delta m_{21}^2=7.5\times 10^{-5}$ $\text{eV}^2$. 

Since current SK solar and KamLAND reactor experiments cannot resolve the tension we studied the potential of the future solar neutrino experiment at HK and the future reactor antineutrino experiment at JUNO to provide a resolution. We find that by combining HK and JUNO data, the SM scenario can be rejected at 7.6$\sigma$ if $\epsilon_{ee}^u=\epsilon_{ee}^d=0.1$. Due to the generalized mass-hierarchy degeneracy, larger negative values of $\epsilon_{ee}^u$ are also allowed at less than 3$\sigma$. We find our conclusions to be robust under reasonable variations of the Earth density profile.
Further, we demonstrated that by combining HK and JUNO data, the SM can be excluded at high confidence for a range of $\epsilon_{ee}^u$ values.

{\it Acknowledgments.} 
We thank S.-H. Seo and M. B. Smy for helpful discussions regarding HK and SK.
KW thanks the University of Hawaii at Manoa for its hospitality during
part of this work. This research was supported in part by the U.S. DOE under
Grant No. DE-SC0010504.



\end{document}